\documentclass[12pt,graphics]{iopart}

\newcommand{\Journal}[4]{{#1} #4 {\bf #2}, #3 }
\usepackage{iopams}  
\usepackage{graphicx}
\usepackage{epsfig}
\usepackage{units}
\usepackage{multicol,graphics,verbatim}
\usepackage{longtable}
\usepackage{colortbl}
\usepackage{color}
\usepackage{rotating}  
\usepackage{calc} 
\usepackage{upgreek}          
\usepackage{ifthen}
\usepackage[colorinlistoftodos]{todonotes}
\usepackage{lineno}

\usepackage{hyperref}

\usepackage{threeparttable}

\newcommand{\LNC}{\em Lett. Nuovo Cimento}

\newcommand{\NPB}{{\em Nucl. Phys.} {\bf B}}

\newcommand{\PLB}{{\em Phys. Lett.}  {\bf B}}
\newcommand{\PRD}{{\em Phys. Rev.} {\bf D}}
\newcommand{\PRC}{{\em Phys. Rev.} C}

\newcommand{\ZPA}{{\em Z. Phys.} A}
\newcommand{\MPA}{{\em Mod. Phys. Lett. A}}

\newcommand{\ARI}{\em Appl. Rad. and Isotopes}

\newcommand{\CPC}{\em Chin. Phys.  C}

\newcommand{\IJMPE}{\em Int. Journal of Modern Physics E}

\newcommand{\JETP}{\em JETP Lett.}

\newcommand{\bpbp}{\mbox{$\beta^+\beta^+$} }
\newcommand{\ecec}{\mbox{$ECEC$} }
\newcommand{\bec}{\mbox{$\beta^+EC$} }

\newcommand{\obb}{0\mbox{$\nu\beta\beta$ decay}} 
\newcommand{\zbb}{2\mbox{$\nu\beta\beta$ decay}}

\newcommand{\nel}{\mbox{$\nu_e$}}

\newcommand{\be}{\begin{equation}}
\newcommand{\ee}{\end{equation}}
\newcommand{\ra}{\rightarrow }
\newcommand{\ema}{\ensuremath{\langle m_{\nu_e} \rangle}~}

\begin{document}
\newcommand{\nuc}[2]{$^{#2}\rm #1$}

\newcommand{\bb}[1]{$\rm #1\nu \beta \beta$}
\newcommand{\bbm}[1]{$\rm #1\nu \beta^- \beta^-$}
\newcommand{\bbp}[1]{$\rm #1\nu \beta^+ \beta^+$}
\newcommand{\bbe}[1]{$\rm #1\nu \rm ECEC$}
\newcommand{\bbep}[1]{$\rm #1\nu \rm EC \beta^+$}

\newcommand{\gerda}{\textsc{Gerda}}
\newcommand{\largeGERDA}{{LArGe}}
\newcommand{\PI}{\mbox{\textsc{Phase\,I}}}
\newcommand{\PIa}{\mbox{\textsc{Phase\,I}a}}
\newcommand{\PIb}{\mbox{\textsc{Phase\,I}b}}
\newcommand{\PIc}{\mbox{\textsc{Phase\,I}c}}
\newcommand{\PII}{\mbox{\textsc{Phase\,II}}}

\newcommand{\geant}{\textsc{GEANT4}}
\newcommand{\mage}{\textsc{MaGe}}

\newcommand{\nPlus}{\mbox{n$^+$ electrode}}
\newcommand{\pPlus}{\mbox{p$^+$ electrode}}

\newcommand{\AOE}{$A/E$}

\newcommand{\order}[1]{\mbox{$\mathcal{O}$(#1)}}

\newcommand{\mul}[1]{\texttt{multiplicity==#1}}

\newcommand{\baseT}[2]{\mbox{$#1\cdot10^{#2}$}}
\newcommand{\baseTsolo}[1]{$10^{#1}$}
\newcommand{\THL}{$T_{\nicefrac{1}{2}}$}

\newcommand{\UBI}{$\rm cts/(kg \cdot yr \cdot keV)$}

\newcommand{\Uflux}{$\rm m^{-2} s^{-1}$}
\newcommand{\Ucpd}{$\rm cts/(kg \cdot d)$}
\newcommand{\Uexpo}{$\rm kg \cdot d$}
\newcommand{\UexpoYear}{$\rm kg \cdot yr$}

\newcommand{\UMWE}{m.w.e.}

\newcommand{\Qbb}{$Q_{\beta\beta}$}

\newcommand{\validate}{\textcolor{blue}{\textit{(validate!!!)}}}

\newcommand{\improve}{\textcolor{blue}{\textit{(improve!!!)}}}

\newcommand{\missing}{\textcolor{red}{\textbf{...!!!...} }}

\newcommand{\quanta}{\textcolor{red}{\textit{(quantitativ?) }}}

\newcommand{\misscite}{\textcolor{red}{[citation!!!]}}

\newcommand{\missref}{\textcolor{red}{[reference!!!]}\ }

\newcommand{\PC}{$N_{\rm peak}$}
\newcommand{\BIC}{$N_{\rm BI}$}
\newcommand{\PAPR}{$R_{\rm p/>p}$}

\newcommand{\PCR}{$R_{\rm peak}$}


\newcommand{\gline}{$\gamma$-line}
\newcommand{\glines}{$\gamma$-lines}

\newcommand{\gray}{$\gamma$-ray}
\newcommand{\grays}{$\gamma$-rays}

\newcommand{\bray}{$\beta$-ray}
\newcommand{\brays}{$\beta$-rays}

\newcommand{\aray}{$\alpha$-ray}
\newcommand{\arays}{$\alpha$-rays}

\newcommand{\betas}{$\beta$'s}


\newcommand{\tab}{{Tab.~}}
\newcommand{\eq}{{Eq.~}}
\newcommand{\fig}{{Fig.~}}
\renewcommand{\sec}{{Sec.~}}
\newcommand{\chap}{{Chap.~}}

 \newcommand{\fn}{\iffalse \fi} 
 \newcommand{\tx}{\iffalse \fi} 
 \newcommand{\txe}{\iffalse \fi} 
 \newcommand{\sr}{\iffalse \fi} 

\today

%
%
%
\title{A search for the radiative neutrinoless double electron capture of $^{58}$Ni}

\author{B. Lehnert$^a$, D. Degering$^b$, A. Frotscher$^a$, T. Michel$^{c}$, K. Zuber$^a$}
\address{$^a$ Institut f\"ur Kern- und Teilchenphysik, Technische Universit\"at Dresden,\\
Zellescher Weg 19, 01069 Dresden, Germany\\
$^b$ VKTA - Strahlenschutz, Analytik \& Entsorgung Rossendorf e.V., 01314 Dresden, Germany\\
$^c$ Erlangen Centre for Astroparticle Physics (ECAP), Friedrich-Alexander-Universit\"at Erlangen-N\"urnberg, 91058 Erlangen, Germany}
\ead{bjoernlehnert@gmail.com}

\begin{abstract}
A search for the radiative neutrinoless double electron capture with single \gray\ emission has been performed in \nuc{Ni}{58}. 
Gamma radiation from a \unit[7286]{g} nickel sample in natural isotope composition was measured for \unit[58.3]{d} with an ultra low background HPGe detector in the Felsenkeller underground laboratory in Dresden, Germany. A new lower half-life limit of \unit[$2.1 \times 10^{21}$]{yr} (\unit[90]{\%} CL) was obtained for this decay mode. This half-life limit is two orders of magnitude higher than the existing limit for \nuc{Ni}{58} and among the best half-life limits for neutrinoless double electron capture decays.

\end{abstract}

\maketitle

\section{Introduction}
\label{intro}

Vast improvements have been made in the field of neutrino physics in recent years. Flavor oscillations in the lepton sector are among the most important discoveries by studying neutrinos 
coming from the sun \cite{sno,superKsolar}, the atmosphere \cite{superKatmos}, high energy 
accelerators beams \cite{minos10,opera10} and nuclear power plants \cite{kamland03}. Lepton flavor violation is explained by neutrino oscillations requiring a non zero neutrino mass; however, no absolute mass scale can be fixed with experiments studying the oscillatory behavior.

Investigating neutrinoless double beta decay (\bb{0}), the decay of a nucleus with a change of atomic number Z by 2 units while leaving
the atomic mass A unchanged, can help to identify total lepton number violation. Additionally it allows to study the absolute neutrino mass scale in combination with endpoint energy studies in single beta decay and cosmological mass bounds \cite{wei13,ade15}.
 
The \obb\ is the gold plated process to distinguish whether neutrinos are Majorana or Dirac particles. The second order weak decay would violate lepton number $L$ by two units and thus is not allowed in the Standard Model (SM):
\be
(Z,A) \ra (Z+2,A) + 2 e^-  \, . 
\ee

This $\Delta L =2$ process can occur in various beyond the standard model theories, however the standard interpretation is the exchange of a light, massive Majorana neutrino. The deduced quantity of such a search, the so-called effective Majorana neutrino mass \ema , is linked with the experimentally observable half-life via
\be
  \label{eq:1}
 \left(T_{1/2}^{0 \nu}\right)^{-1} = G^{0 \nu} \left| M^{0\nu}\right|^2 \left(\frac{\ema}{m_e}\right)^2 \, ,
\ee
where $G^{0 \nu}$ is the phase space factor and $M^{0\nu}$ describes the nuclear transition matrix element.
The effective Majorana neutrino mass is given by $\ema = \left| \sum_{i}U_{ei}^2m_i\right|$ with $U_{ei}$ being the corresponding  elements in the PMNS mixing matrix extended with two Majorana CP-phases.
In addition, the SM process of neutrino accompanied double beta decay,
\be
(Z,A) \ra (Z+2,A) + 2 e^- + 2 \bar{\nu}_e \quad (\zbb) \,
\ee

can be investigated, which is expected and observed with half-lives between \unit[$10^{18}$ and $10^{24}$]{yr}. For recent reviews see \cite{avi08,Rodejohann:2011fr,Rodejohann:2012cc}.

While a lot of activities are ongoing on the neutron rich side of the isobars, the interest in the proton rich side only increased recently.
Here the alternative processes of electron capture (EC) in combination with positron decay can occur. Three different decay modes are distinguished:
\begin{eqnarray}
(Z,A) &\ra (Z-2,A) + 2 e^+ + (2 \nel)  \hspace{2pc}& \mbox{(\bpbp{})}\\
(Z,A) + e^- &\ra (Z-2,A) + e^+ + (2 \nel)  \hspace{2pc}&\mbox{(\bec{})}\\
(Z,A) + 2 e^-&\ra (Z-2,A) + (2 \nel)  \hspace{2pc}&\mbox{(\ecec{})}
\end{eqnarray}

Decay modes containing an EC emit X-rays or Auger electrons (from now on only called X-rays) created during filling of the atomic shell vacancy in the daughter atom. 
Decay modes containing a positron emission have a reduced Q-value as each generated positron accounts for a reduction 
of 2 $m_ec^2$ in the phase space.
Thus, the largest phase space is available in the ECEC mode and makes it the most probable one. However, the \bbe{2} is also the most difficult to detect, only producing two X-rays and neutrinos in the final state instead of \unit[511]{keV} \grays\
resulting from the decay modes involving positrons. Even more, being a rare process this could be disturbed by 
more subtile effects as described for another system in \cite{mic14}. For the case studied here \nuc{Ni}{58}
could make a thermal neutron capture ($\sigma \approx$ 4.6 b) to \nuc{Ni}{59}, which can decay by ordinary EC.
This decay can cause a second K-shell fluorescence with a probability of about 10$^{-4}$ thus producing a similar signature as 2$\nu$ECEC. 
Furthermore, it has been shown that neutrinoless $\beta^+$EC transitions have an enhanced sensitivity to right-handed weak
currents (V+A interactions) \cite{hir94} and thus would help to disentangle the physics mechanism of \obb, if observed.

In the \bbe{0} mode there are only two X-rays and if there is no other particle in the final state this would violate energy and momentum conservation. It was suggested that the energy is released radiatively as a single internal radiative Bremsstrahlung \gray, a double \gray\ or an $e^-e^+$-pair \cite{Vergados:1983do,Doi:1993hj}. The \bbe{0} decay is generally expected to have much larger half-lives than the $0\nu\beta^-\beta^-$ decay for the same effective neutrino mass. One of the reasons is the additional coupling factor $\alpha \approx 1/137$ 
required for the additional particles. The radiative single gamma decay mode of \bbe{0} is experimentally the easiest to access and is the focus of this search.

\begin{figure}[ht]
\centering
\includegraphics[width=0.9\textwidth]{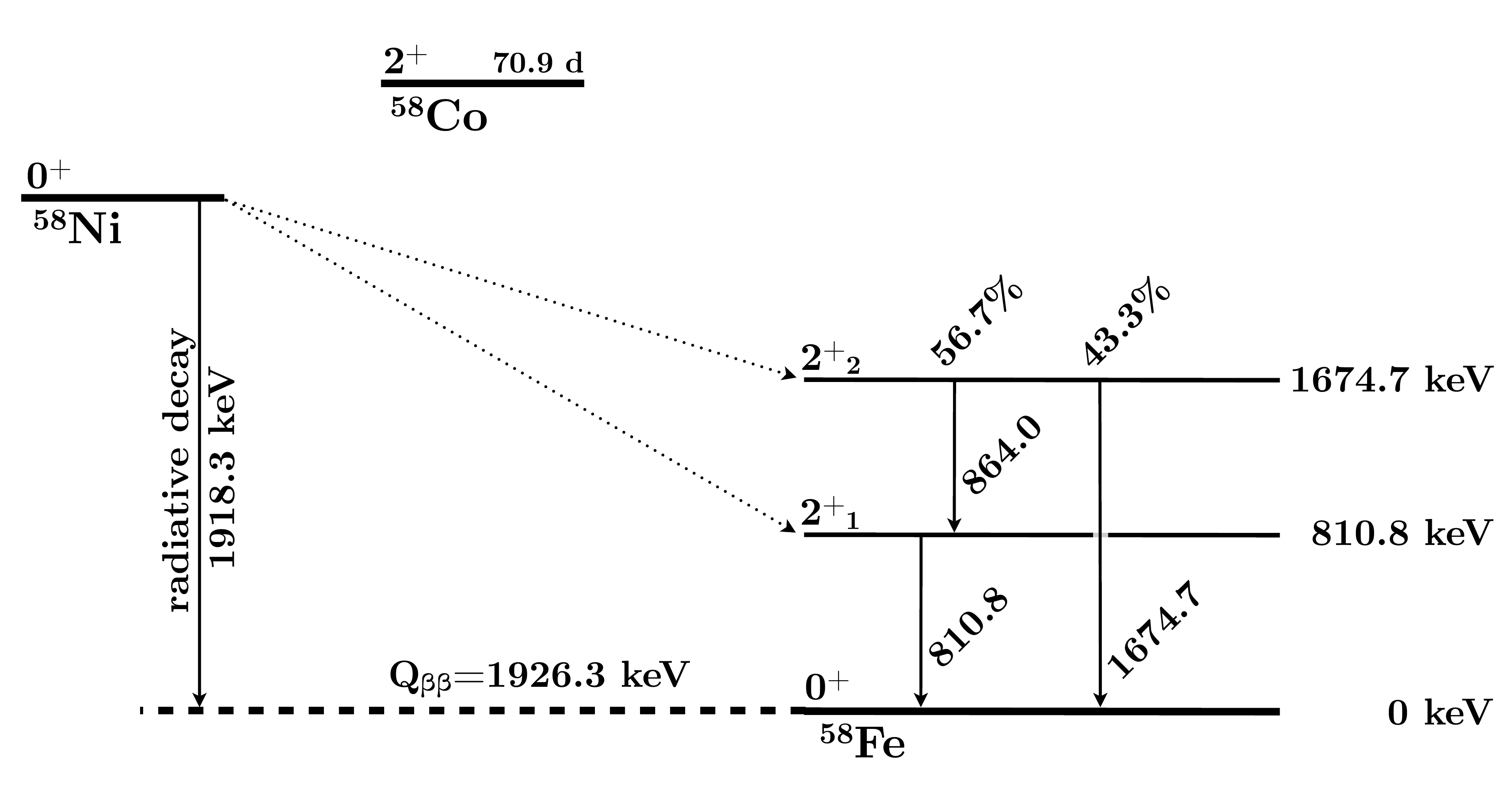}
\caption{Level scheme of the decay \nuc{Ni}{58} - \nuc{Fe}{58} with illustration of the radiative decay. The excited state transitions are not investigated in this work.}
\label{fig:level}
\end{figure}

\nuc{Ni}{58} is a favorable isotope for investigating \bbe{0} having one of the largest natural abundance among all double beta emitters of \unit[68.08]{\%}. Natural nickel is comparably cheap and easy to handle. The large Q-value of \unit[($1926.3 \pm 0.7$)]{keV}  \cite{ame12} allows for decays into two excited states of the daughter nuclide \nuc{Fe}{58} and for the \bbep{2/0} mode. The decay scheme of \nuc{Ni}{58} into \nuc{Fe}{58} is illustrated in \fig \ref{fig:level}. The following decay modes are feasible to investigate with gamma spectroscopy:
\begin{eqnarray}
2\nu {\rm ECEC}:\ & ^{58}{\rm Ni} + 2 e^- &\rightarrow\ ^{58}{\rm Fe}(2^+_1) +2\nu_e  + 2\gamma_{\rm shell}  + \gamma_{(810.8~\rm keV)} \\
2\nu {\rm ECEC}:\ & ^{58}{\rm Ni} + 2 e^- &\rightarrow\ ^{58}{\rm Fe}(2^+_2)+2\nu_e  + 2\gamma_{\rm shell} + \gamma_{(1674.8~\rm keV)} \\
 & &\rightarrow\ ^{58}{\rm Fe}(2^+_2)+2\nu_e  + 2\gamma_{\rm shell} + \gamma_{(810.8~\rm keV)} + \gamma_{(864.0~\rm keV)}  \nonumber \\
2\nu {\rm EC}\beta^+:\ & ^{58}{\rm Ni} + e^- &\rightarrow\ ^{58}{\rm Fe}(g.s.)+2\nu_e +e^{+} + \gamma_{\rm shell}  + 2\gamma_{(511~\rm keV)}\\
0\nu {\rm ECEC}:\ & ^{58}{\rm Ni} + 2 e^- &\rightarrow\ ^{58}{\rm Fe}(g.s.)  + \gamma_K + \gamma_L + \gamma_{(1918.3~\rm keV)}\ \, \label{eq:0necec}
\end{eqnarray}

with shell de-excitations ($\gamma_{\rm shell}$, $\gamma_K$, $\gamma_L$) also possible as Auger electrons as mentioned.
The first two decay modes are excited state transitions where the search is based on the de-excitation \grays. In principle here the neutrino-less mode is also accessible with an additional radiative release of the remaining energy e.g.\ in form of a single Bremsstrahlung \gray. 
The experimental signature of the \bbep{2} mode are the two \unit[511]{keV} annihilation \grays; however, this \gline\ is also created by various background processes and cannot be distinguished from the \bbep{2} process unless the X-rays are detected in coincidence. 
Due to contaminations in the measured sample, excited state searches were not feasible (see section \ref{analysis}). Hence, the search in this paper considers only the radiative \bbe{0} decay (\eq \ref{eq:0necec}). The expected Bremsstrahlungs \gray\ has an energy of \unit[$(1918.3\pm0.7)$]{keV} and accounts together with the electron binding energies of the K shell (\unit[7.11]{keV}) and L shell (\unit[0.85]{keV}) \cite{TOI96} in \nuc{Fe}{58} for the Q-value of the decay. The uncertainty on the Bremsstrahlungs \gray\ energy is dominated by the Q-value. Possible small energy shifts due to effects in the atomic shell are well covered by this uncertainty. In principle one of the electron captures could also occur from the M shell which is however suppressed by a smaller overlap of the s-orbital with the nucleus in higher shells.

\begin{table}[htbp]
\begin{center}
\begin{tabular}{|c|c|c|}
\hline
Decay mode & $T_{1/2}$ &  Reference\\
\hline
\bbep{0/2} 0$_{\rm g.s.}^+$            & $> 7 \times 10^{20}$ (\unit[68]{\%} CL) & 1993 \cite{vas93}\\
\bbep{0/2} 2$_1^+$ (\unit[810.8]{keV}) & $> 4 \times 10^{20}$ (\unit[68]{\%} CL) & 1993 \cite{vas93}\\
\bbe{0/2} 2$_2^+$ (\unit[1674.7]{keV})  & $> 4 \times 10^{19}$ (\unit[90]{\%} CL) & 1982 \cite{bel82}\\
\bbe{0} 0$_{\rm g.s.}^+$ + 2$_1^+$ (\unit[810.8]{keV})& $> 2.1 \times 10^{19}$ (\unit[68]{\%} CL) & 1984 \cite{Norman:1984cq} \\
\hline
\end{tabular}
\medskip
\caption{\label{t:nimessungen} Currently known half-life limits on various \nuc{Ni}{58} decay modes. 
The $\gamma$-line energies for the transitions are given in brackets.
}
\end{center}
\end{table}

Searches on some of the decay channels of \nuc{Ni}{58} so far have been performed by  \cite{bel82,vas93,Norman:1984cq} and are compiled in Tab.~\ref{t:nimessungen}. The searches were mostly designed for coincidences of multiple \grays. The limit in Ref.~\cite{Norman:1984cq} includes, among others, the radiative \bbe{0} decay mode of \nuc{Ni}{58}. It is the only previous result for this transition.

There exist a theoretical predictions for the radiative \bbe{0} process in \nuc{Ni}{58} with a half-life estimation between \unit[\baseT{2}{35} and \baseT{3}{36}]{yr} for an effective Majorana neutrino mass of \unit[1]{eV}  \cite{Merle:2009}.
%
%
A potential resonance enhancement for \bbe{0} is expected for isotopes in which an excited state of the daughter nuclide is energetically degenerated with the ground state of the mother. This may lead to several orders of magnitude faster rates \cite{Sujkowski:2004dub,ber83,Kotila:2014ira}. Precision mass measurements using Penning traps have been performed to search for this effect, see for example
\cite{Eliseev:2011}. However, such a situation is not realized in \nuc{Ni}{58} and thus this search focuses on the radiative transition
only.\\ 



\section{Experimental Setup}
\label{setup}
\begin{figure}[ht]
\begin{center}
  \includegraphics[width=0.8\textwidth]{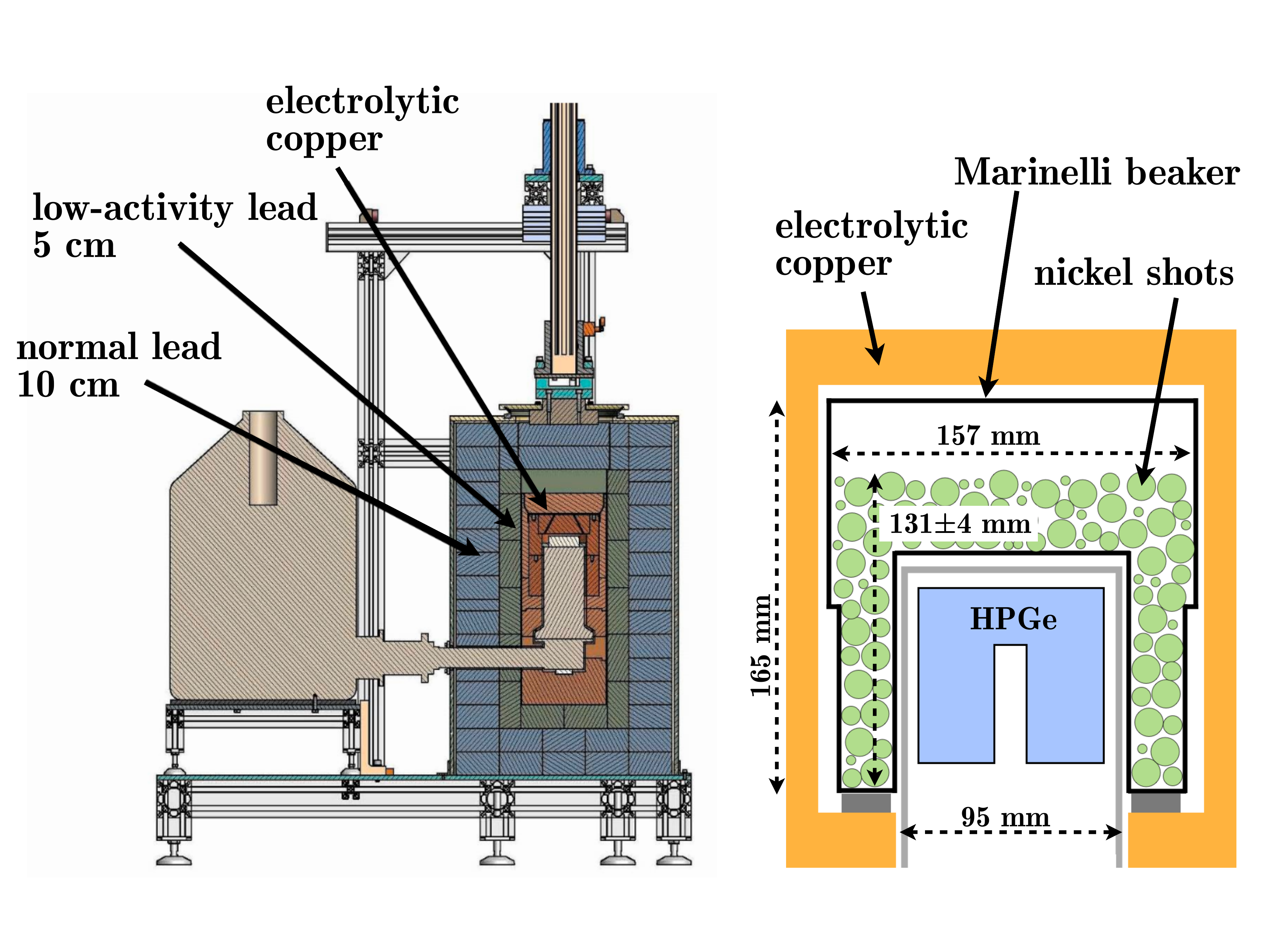}
\caption{\small Left: Schematic drawing of the low background detector setup. Right: Nickel-shots in a Marinelli beaker on the HPGe detector.}
\label{fig:DetectorAndSetup}
\end{center}
\end{figure}

The measurement was performed in the Felsenkeller Underground Laboratory in Dresden, Germany, with a shielding of \unit[110]{m.w.e.} overburden reducing the muon flux to \unit[\baseT{0.6}{-3}]{cm$^{-2}$s$^{-1}$} \cite{Niese98,FelixDPG}. A sample of \unit[7286]{g} of nickel shots was acquired from Alfa Aesar with a guaranteed purity of \unit[$>$99.95]{\%}. The granular size of the shots varies between \unit[3]{mm} and \unit[25]{mm}. They are filled into a standard \unit[1.5]{l} Marinelli beaker with an inner recess fitting onto the end cap of the HPGe detector. A schematic drawing of the arrangement can be seen in \fig \ref{fig:DetectorAndSetup}. The effective density of the nickel shot packing was determined to be \unit[($5.18\pm0.21)$]{g/cm$^3$} compared to \unit[8.908]{g/cm$^3$} of solid nickel.

The sample was positioned on an ultra low background HPGe detector with a relative efficiency of \unit[90]{\%} routinely used for low background gamma spectroscopic measurements. It is surrounded by a \unit[5]{cm} copper shielding embedded in another shielding of \unit[15]{cm} of clean lead. The inner \unit[5]{cm} of the lead shielding have a specific activity of \unit[($2.7 \pm 0.6$)]{Bq/kg}  $^{210}$Pb  while the outer \unit[10]{cm} have \unit[($33 \pm 0.4$)]{Bq/kg}. The spectrometer is located in a measuring chamber which acts as an additional shielding.
Furthermore, the detector is constantly held in a nitrogen atmosphere to avoid radon. The data were collected with a 8192 channel MCA from ORTEC recording energies up to \unit[2.8]{MeV}. More details can be found in \cite{deg09,degering08}. 

\begin{figure}[ht]
\begin{center}
  \includegraphics[width=0.99\textwidth]{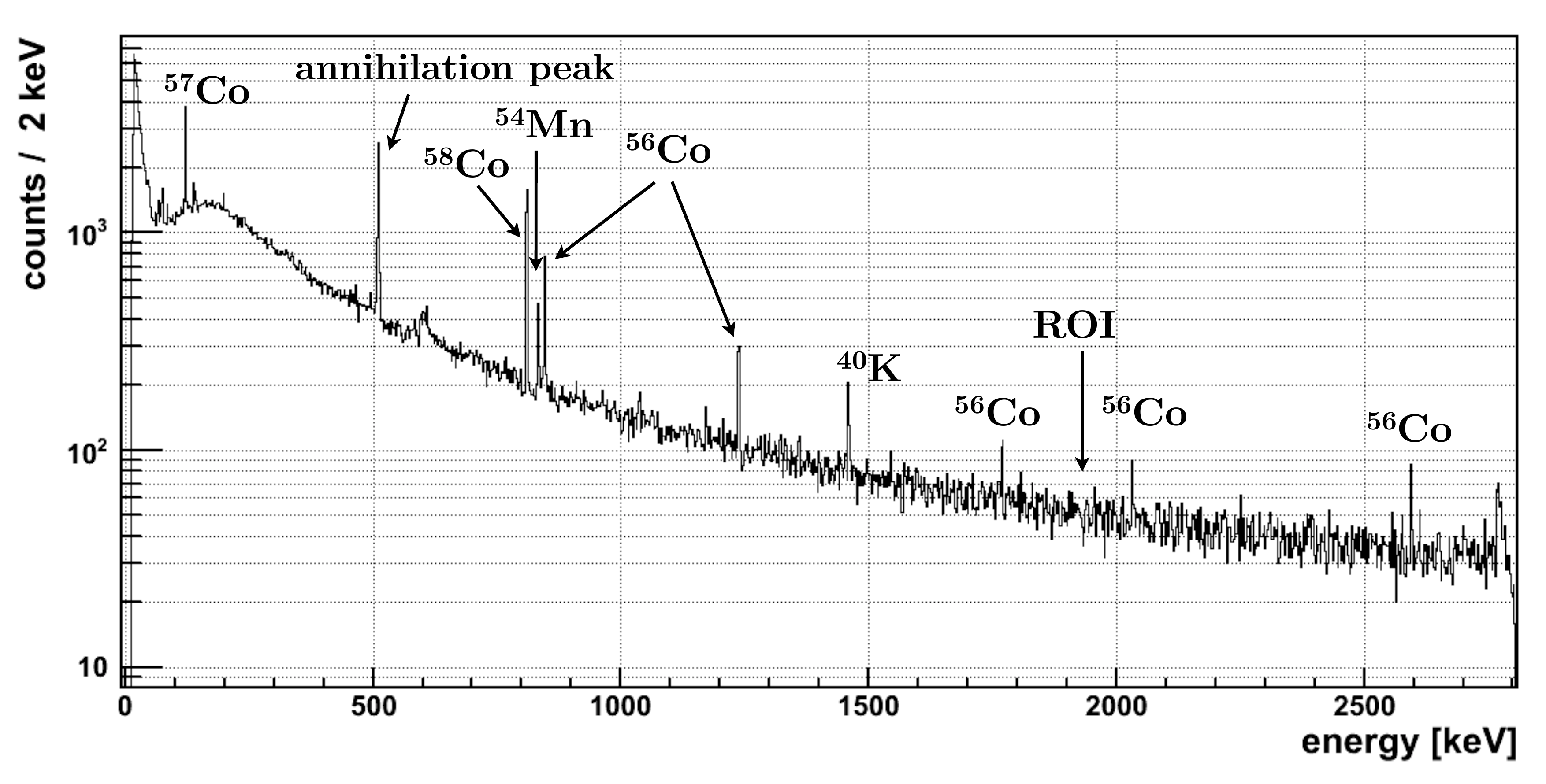}
\caption{\small Energy spectrum of the nickel sample after the full measurement period of \unit[58.3]{d}. }
\label{fig:FullSpec}
\end{center}
\end{figure}

The nickel sample was stored underground for about two weeks before the measurement. The previous history of exposure to cosmic radiation is unknown; however a clear signal of cosmogenic produced \nuc{Co}{56} (\unit[\THL=77.3]{d}, \unit[846.8]{keV}, \unit[1238.3]{keV}, \unit[1771.4]{keV}, \unit[2034.8]{keV} and \unit[2598.5]{keV} \gline), \nuc{Co}{57} (\unit[\THL=271.8]{d}, \unit[122.1]{keV} and  \unit[136.5]{keV} \gline), \nuc{Co}{58} (\unit[\THL=70.9]{d}, \unit[810.8]{keV} \gline) and \nuc{Mn}{54} (\unit[\THL=312.3]{d}, \unit[834.9]{keV} \gline) could be observed (see \fig \ref{fig:FullSpec}). The decay of \nuc{Co}{58} populates the same excited states in  \nuc{Fe}{58} as the \bbe{0/2} \nuc{Ni}{58} decay. Due to this non-reducible background, the excited state transitions in \nuc{Ni}{58} could not be investigated.

The efficiency calibration was performed by a mixture of analytically pure SiO$_2$, the reference materials RGU and RGTh from IAEA \cite{RGU} and KCl. This activity standard contained specific activities of \unit[($107.0\pm0.3$)]{Bq/kg} \nuc{U}{238}, \unit[($113.0\pm1.2$)]{Bq/kg} \nuc{Th}{232} and \unit[($106.9\pm2.1$)]{Bq/kg} \nuc{K}{40}. It was filled in the same container type as used for the nickel sample with the same measuring geometry. Thus, calibration source and measuring sample differ only in their self absorption behavior.


The stability of energy calibration and resolution over the entire \unit[58.3]{d} measuring time was checked by the \glines\ of the cosmogenic cobalt isotopes in the nickel sample. It was shown that no difference in energy resolution occurred between the long term measurement and the short term calibration.

\section{Analysis and Results}
\label{analysis}
\begin{figure}[ht]
\begin{center}
 \includegraphics[width=0.99\textwidth]{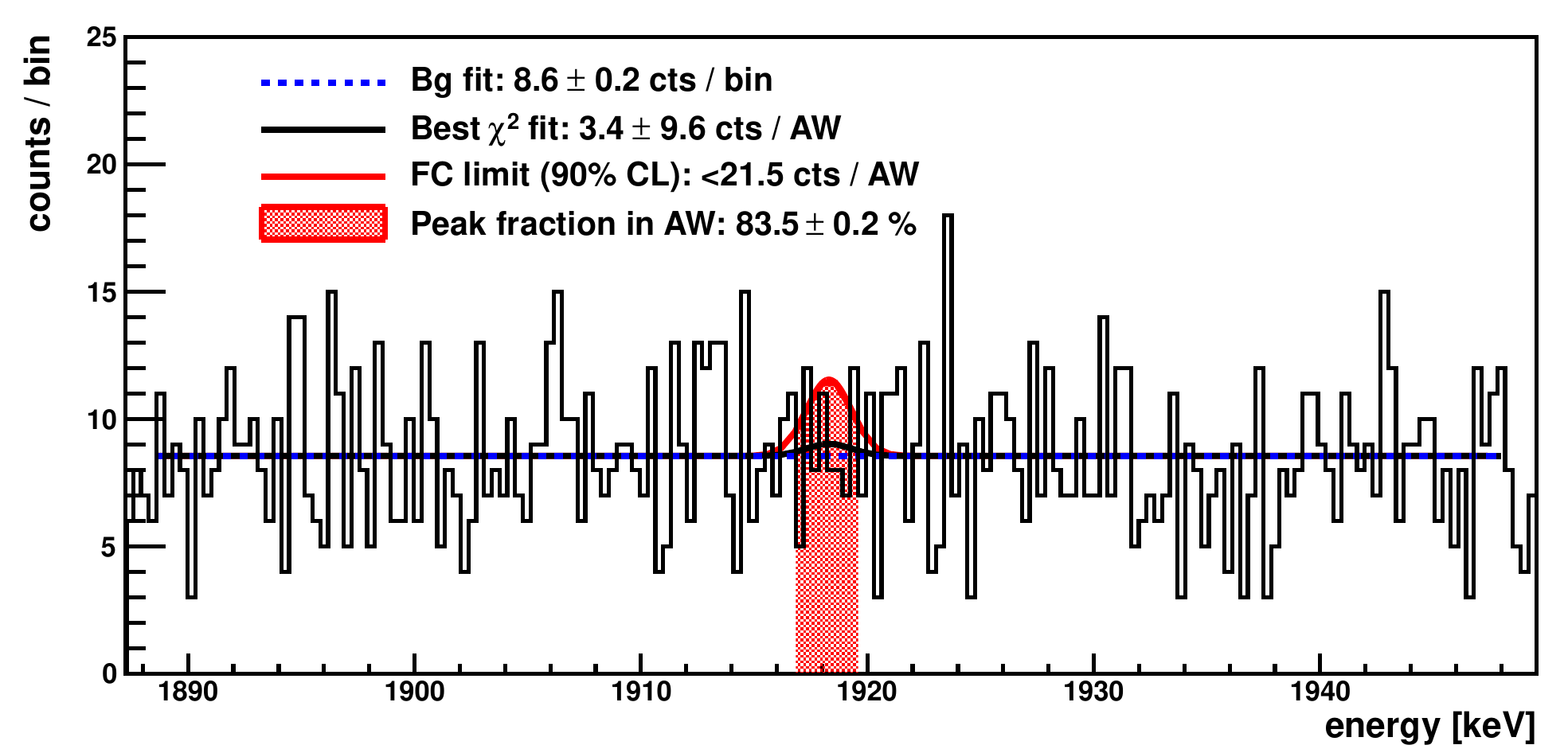}
\caption{\small Background region, analysis window (AW) shown in shaded red and maximum likelihood fit shown in black. Also shown is the background fit in blue and the peak with \unit[90]{\%} CL Feldman Cousins count limit in red. The plot shows all analysis parameters at their nominal values.}
\label{fig:ni58}
\end{center}
\end{figure}

A total exposure of \unit[1.16]{$\rm kg \cdot yr$} was accumulated.  
The analysis is purely based on the search for a \gline\ in a peak region centered around the expected energy of the Bremsstrahlung photon of \unit[1918.3]{keV}. There are no prominent \glines\ around this region as can be seen in the spectrum shown in \fig \ref{fig:ni58}. The background is dominated by the residual muon flux in the underground lab. 

The energy resolution at the \gline\ energy was taken from the calibration as \unit[($2.33 \pm 0.01$)]{keV} FWHM or \unit[0.12]{\%}. 

The full energy detection efficiency of a \unit[1918.3]{keV} \gray\ from the nickel sample was determined with Monte Carlo (MC) simulations based on Geant4. The precise detector geometry was implemented in the code framework MaGe \cite{Boswell:hc} developed for particle propagation at low energies. The code was validated with the calibration source in the same geometry as the nickel sample. For \glines\ above \unit[1]{MeV} the relative difference of detection efficiency between MC and calibration measurement was on average \unit[5]{\%} but never more than 10\%. The full energy detection efficiency at \unit[1918.3]{keV} for the nickel sample was determined as \unit[($1.21\pm0.07$)]{\%} with the uncertainty based on the statistical and systematic uncertainties in the MC simulations and the uncertainty arising from the effective density of the nickel sample.

No evident peak was observed at \unit[1918.3]{keV}. An unbound maximum likelihood fit of a Gaussian peak with a constant background resulted in $3.4\pm9.6$ counts. In the following a limit on the signal counts is set using the Feldman Cousins method \cite{{fc98}}. This method combines the advantages of yielding non-negative signal count limits in all cases and providing a natural transition between a one-side and a two-sided confidence interval resulting in better coverage for this region. 

For the limit setting, the peak region is combined into a single analysis window of \unit[2.75]{keV} (8 MCA bins) which covers the Gaussian peak with \unit[$83.5$]{\%}. The constant background level was estimated in a \unit[$\pm30$]{keV} region left and right of the peak excluding the peak region by \unit[$\pm5$]{keV}. No visible \glines\ were found in the background region. A potential background line at \unit[1919.5]{keV} from \nuc{Ni}{57} could be present. However, its short half-life of 36 hrs and the underground storage time of 2 weeks before the measurement will remove this contribution. A potential in-situ production of \nuc{Ni}{57} via (n,2n) reactions on \nuc{Ni}{58} is unlikely due to the shielding. This is supported by the non-observation of the \unit[1377.6]{keV} line of \nuc{Ni}{57} which has a more than
5 times higher intensity and higher detection efficiency. The expected background in the analysis window is 68.5 events corresponding to \unit[21.5]{cts/(keV$\cdot$kg$\cdot$yr)}. 71 events are observed and the \unit[90]{\%} CL upper count limit based on Feldman Cousins is 21.5 event. The analysis window (red area) along with the background region is shown in \fig \ref{fig:ni58}. 

The lower half-life limit was determined including uncertainties on the background counts, the detection efficiency, the energy resolution and the peak position. The lower half-life limit of \baseTsolo{5} MC experiments was calculated changing the input parameters of the analysis randomly. The background counts were varied according to the Poisson distribution of the expectation value. The efficiency, energy resolution and peak position were varied according to Gaussian distributions with the width of their uncertainties. The median of the limit distribution was taken as the lower half-life limit including systematic uncertainties and found to be 
$$T_{1/2}>2.1\cdot10^{21}\, {\rm yr}\hspace{2pc} (90\,\%\ {\rm C.L.})\, .$$

The analysis window shown in \fig \ref{fig:ni58} is one realization the MC experiments in which all input parameters are at their nominal values. A sensitivity study was performed along the same principle. Here, the observed number of events in the analysis window was additionally randomized according to the background only hypothesis, i.e.\ according to the Poisson distribution of the background expectation in the analysis window. The corresponding sensitivity of the experiment is \unit[\baseT{4.1}{21}]{yr}, taken as the median of the sampled distribution. The sensitivity is larger than the derived limit due to an upward fluctuation of background in the peak region.

\section{Summary}
\label{conclusion}
The radiative neutrinoless double electron capture with single \gray\ emission has been
investigated in \nuc{Ni}{58}. No signal has been observed and a lower half-life limit was set to \unit[$2.1 \times 10^{21}$]{yr} (\unit[90]{\%} CL). This limit is two orders of magnitude stronger than the previous limit of \cite{Norman:1984cq}. \nuc{Ni}{58} is a promising isotope to investigate this transition having one of the largest natural abundance of all double beta isotopes, having a large Q-value and being reasonable cheap and easy to handle. A further measurement deeper underground is envisaged.

\section*{Acknowledgement}
This work is supported by the Deutsche Forschungsgemeinschaft DFG (grant ZU123/7-1,MI1507/1-1).

\end{document}